\documentstyle[12pt,epsf]{article}
\newcommand{\be} {\begin{equation}}
\newcommand{\ee} {\end{equation}}
\newcommand{\bdm} {\begin{displaymath}}
\newcommand{\edm} {\end{displaymath}}
\newcommand{\bc} {\begin{center}}
\newcommand{\ec} {\end{center}}
\newcommand{\beqa} {\begin{eqnarray}}
\newcommand{\eeqa} {\end{eqnarray}}
\newcommand{\nn} {\nonumber}
\newcommand{\ra} {\rightarrow}

\newcommand{\bbf}{\bf}
\begin{document}
\sloppy
\bc
{\bf\Large $\gamma^* \gamma^*$ Reactions at High Energies}
\ec
\vskip 1truecm
\bc
A. Donnachie\\
Department of Physics and Astronomy,
University of Manchester\\Manchester M13 9PL, UK\\
email: {\tt ad@a3.ph.man.ac.uk}
\ec
\bc
H.G.Dosch\\
Institut f\"ur Theoretische Physik der Universit\"at Heidelberg\\
Philosophenweg 16, D-69120 Heidelberg\\
email: {\tt h.g.dosch@thphys.uni-heidelberg.de}
\ec
\bc
M. Rueter\footnote{Supported by a MINERVA-fellowship}\\
School of Physics and Astronomy\\
Department of High Energy Physics, Tel Aviv University\\
69978 Tel Aviv, Israel\\
email: {\tt rueter@post.tau.ac.il}
\ec
\vskip 2truecm
\bc
{\bf Abstract}
\ec
\medskip
The total hadronic $\gamma^{*}\gamma^{*}$ cross sections at high energy are 
calculated as a function of energy and photon virtuality in a model combining
Reggeon exchange, the quark box diagram (a fixed pole in Regge language) and
soft and hard pomeron exchanges evaluated in the context of dipole-dipole
scattering. Good agreement is obtained with the data for the real 
$\gamma\gamma$ cross section and for the real photon structure function
$F_2^\gamma$. However the model prediction for the $\gamma^{*}\gamma^{*}$
cross section is too small. This is attributed to an incorrect 
extrapolation of the $Q^2$ dependence of the hard pomeron adopted here.
Parametrising it independently shows that the hard
part of the cross section can be well represented by a simple Regge pole with
intercept $\sim 1.3$. \newpage

\section{Introduction}

The energy available for $\gamma^{(*)}\gamma^{(*)}$ physics at LEP2 is
opening a new window on the study of diffractive phenomena, both 
non-perturbative and perturbative. These phenomena occur in each of 
untagged, single-tagged and double-tagged reactions via the total
hadronic $\gamma\gamma$ cross section, $\sigma_{\gamma\gamma}$; the 
structure function of the real photon, $F_2^{\gamma}$ (or equivalently 
the $\gamma^{*}\gamma$ cross section); and the total hadronic $\gamma^{*}
\gamma^{*}$ cross section, $\sigma_{\gamma^*\gamma^*}$ respectively.
Thus in principle it is possible to study diffraction continuously
from the quasi-hadronic regime dominated by non-perturbative physics
to the realm of perturbative QCD with either single or double hard
scales. Although measurement of $\sigma_{\gamma^{*}\gamma^{*}}$ at high
energies breaks new ground, data on $F_2^{\gamma}$ at small $x$ is equally
significant. There has been considerable effort recently to understand the 
solution to the BFKL equation \cite{bfkl} in order to find the correct 
interpretation of the proton structure function at small $x$ and large $Q^2$. 
As this has implications directly for $F_2^{\gamma}$ at small $x$, and 
ultimately for $\sigma_{\gamma^{*}\gamma^{*}}$, it is worth summarising 
the current situation.

The leading order (LO) BFKL resummation \cite{bfkl} of the flavour singlet
evolution equations was initially thought to provide a powerful tool for
understanding the small-$x$ limit of the proton structure function. It
predicted the ``hard pomeron'', the structure function behaving as
$\sim x^{-\lambda}$ with $\lambda \sim 0.5$ in conformity with the rapid
rise with decreasing $x$ discovered at HERA \cite{h1,zeus}. However when 
the NLO corrections to the BFKL resummation were calculated \cite{fl,cc1}
the highest eigenvalue of the BFKL equation was found to be negative and
large, so much so that it could lead to negative values of $\lambda$. It
has been pointed out \cite{s,cc2} that this problem can be alleviated by
identifying that part of the NLO corrections with double logarithms in 
the transverse momenta and resumming them. After ensuring the consistency
of these double logarithms to all orders the perturbation series is much 
more convergent. The result in \cite{cc2} is particularly stable and
gives $\lambda$ in the range 0.26 to 0.32 for the HERA kinematic regime.
The double logarithms are closely associated with the choice of scale.
It has been stressed \cite{bfklp} that the NLO corrections must necessarily
contain both renormaliztion and scale ambiguities, and shown that the NLO
corrections to the BFKL are controllable if appropriate renormalization 
scales and schemes are used, specifically the BLM \cite{blm} scheme for 
scale setting. In this approach the intercept of the NLO BFKL retains an 
extremely weak dependence on $Q^2$ and is smaller than the original
BFKL intercept, having $\lambda \sim 0.2$ over the relevant experimental 
$Q^2$ range. 

In the standard application of the DGLAP evolution equations \cite{dglap}
the rapid rise of the proton structure function at small $x$ is associated
with a singularity at $N = 0$ in the Mellin transform of the DGLAP splitting
function. This singularity is not apparent in the original BFKL LO summation
\cite{bfkl} nor in the NLO corrections of \cite{s,cc2}. It has been argued 
\cite{cdl} that by analytically continuing in $Q^2$ one can conclude that
the singularities in the complex $N$-plane of the Mellin transform of the
proton structure function must also be present at small $Q^2$, and the
perturbative evolution cannot generate new singularities that appear only
at high $Q^2$. In this picture it is natural to associate the rapid rise of 
the proton structure function at small $x$ with a second pomeron \cite{dl},
very much in the spirit of the BFKL approach. This hypothesis was successfully
tested in \cite{dl}, assuming that the contribution to the proton structure 
function from branch points is much weaker than that from poles and including 
only three of the latter: the standard reggeon and soft-pomeron exchanges 
known from purely hadronic interactions and a new hard pomeron with intercept 
$1 + \epsilon_0$. An excellent fit was obtained to the data for $x < 0.07$ 
and $0 \le Q^2 \le 2000$ GeV$^2$, giving $\epsilon_0 = 0.42$.

In references \cite{bfkl} and \cite{fl} through \cite{bfklp} the running of 
the coupling constant was not taken fully into account. A systematic approach 
to the BFKL equation at NLO with running coupling is presented in \cite{t},
adopting the BLM \cite{blm} scale-setting procedure. The effect is quite
dramatic, removing all singularities to the right of $N = 0$ in the complex 
$N$-plane of the Mellin transform of the proton structure function i.e. there
is no power-like behaviour at small-$x$ from perturbative evolution. The 
solution factorizes into a part describing the evolution in $Q^2$ and a 
part describing the input distribution which is infrared dominated and 
non-calculable. Thus the BFKL equation can predict only the evolution in 
$Q^2$ of the structure function, the $x$ dependence at small $x$ being given 
partly by the evolution and partly by the input distribution. The evolution 
at small $x$ differs significantly from that predicted from a standard NLO 
DGLAP treatment. A global fit to the proton structure function is very 
successful.

Thus we are currently in the position of having several apparently disparate 
views of small-$x$ physics, and the quality of the corresponding fits to the 
$F_2(x,Q^2)$ data are such that the latter do not provide the necessary 
discrimination. Additional reactions are required, for example measurement
of the proton longitudinal structure function $F_L(x,Q^2)$ or $\sigma_
{\gamma^*\gamma^*}$ at high energies. We concentrate on the latter,
with emphasis on the information which may be obtained from LEP. The
advantage of $\gamma^* \gamma^*$ interactions is the absence of an 
initial non-perturbative state (e.g. the proton) and the presence of
a hard component in the photon wave function, even for the real photon.
Together these ensure that the ``hard pomeron'' plays a decisive r\^ole
even at small virtualities. This has been demonstrated recently \cite{ddr} 
for the real $\gamma\gamma$ cross section, the real photon structure function
$F_2^\gamma$, and the reaction $\gamma^{*}\gamma^{*} \rightarrow V_1 V_2$,
where $V_1$, $V_2$ are any one of $\rho$, $\omega$, $\phi$, $J/\psi$. 

The application of the BFKL formalism to $\gamma^{*}\gamma^{*}$ has been
considerd by Brodsky et al \cite{bhs} and by Boonekamp et al \cite{bdrw}.
In the BFKL formalism there is a problem at LLO in setting the two mass
scales on which the cross section depends: the mass $\mu^2$ at which the 
strong coupling $\alpha_s$ is evaluated and the mass $Q_s^2$ which provides 
the scale for the high energy logarithms. Brodsky et al \cite{bhs} argue
that $\mu^2 \sim 10^{-1}Q_1Q_2$ and that $Q_s^2 \sim 10^2Q_1Q_2$ are 
reasonable choices. However the result is very sensitive to these parameters
and by way of illustration they show that changing $\mu^2 \ra 4\mu^2$ or
$Q_s^2 \ra Q_s^2/4$ alters the predicted cross section by factors of 
$\sim 1/4$ or $\sim 4$ respectively in a typical LEP experiment. In an
attempt to overcome the scale problem, Boonekamp et al \cite{bdrw} take a
phenomenological approach to estimate the NNLO effects, making use of a fit
\cite{bfklf} to the proton sctructure function using the QCD dipole picture 
of BFKL dynamics. This reduces both the size of the BFKL cross scetion and 
its energy dependence.

The objective of this paper is to provide a realistic estimate of the known 
part of the cross section i.e. everything but the ``hard pomeron'' component 
although an estimate of that will also be given. There is a natural division 
of the contributions to the total cross section into quark-antiquark and 
multiple-gluon exchange in the $t$-channel. In terms of Regge-language the
quark-antiquark exchange corresponds to that of a Reggeon and the gluon 
exchange corresponds to that of the Pomeron. In the language of structure 
functions the Reggeon exchange is mostly the valence quark contribution, the 
Pomeron exchange mostly the gluon contribution. Due to the different reference
frames of the two approaches there is no strict one-to-one correspondence. 
The Regge language is applicable to photon-photon scattering for values 
of $x$ sufficiently small, say $x \leq 0.1$, just as for deep inelastic
scattering on nucleons. There is an important difference between 
hadron-hadron and photon-hadron or photon-photon scattering. To lowest order 
in QED there is no quadratic unitarity relation for the scattering amplitude 
and hence fixed singularities in the complex $J$-plane are possible and at 
least in one case (Compton scattering on hadrons) are required \cite{BGLL67a} 
- \cite{Sin67} by current algebra relations\cite{Fub66}. However this term
is purely real and so does not contribute to the nucleon structure function
$F_2$. In contrast, the box diagram in photon-photon scattering, which
received much attention especially in the pre- and early QCD area (see 
\cite{Fey69} - \cite{HR79} and the literature quoted therein) gives rise to 
a fixed $J$-plane singularity which does contribute to $F_2^\gamma$. This is
an important point as it means that the valence-quark term and the box
diagram must {\it both} be included as they correspond to different $J$-plane
singularities.

It has been rather usual to approximate the soft-pomeron contribution to
$\gamma^{(*)}\gamma^{(*)}$ scattering by assuming dominance of the vector 
meson resonances in the photon-channels. This is formally correct if all 
resonances are taken into account but that is impractical. Taking only one 
or two resonances into account can be misleading even at small virtuality, 
and furthermore the treatment of the longitudinal polarization of the photon 
is rather arbitrary. We therefore adopt an approach which considers the 
hadronic part of the photons as a superposition of quark-antiquark dipoles 
rather than vector mesons. In model investigations \cite{DGP98} it has been 
shown that this approach is much more economical since even for real photons 
the free photon wave function with a suitably chosen quark mass gives a more 
realistic description of a confined system than a superposition of many vector
meson states. It is especially suited for treating the hard part of the 
photon, which in vector dominance can only be described by a superposition 
of infinitely many resonances.  We thus expect, and obtain, large deviations 
from VMD. 

The soft pomeron will be described by the interaction of the dipoles with the 
physical vacuum which has led to a satisfactory quantitative description of
hadron-hadron scattering. For small dipole sizes its coupling is proportional 
to the product of the squares of the dipole radii and therefore is strongly 
suppressed for scattering of photons with high virtuality. The coupling of 
two perturbative gluons to small dipoles has a very different dependence on
the radii which can be rather well described by:
\bdm
\frac{R_1^2 R_2^2}{R_1^2+R_2^2}.
\edm
Hence we anticipate, not unexpectedly, a strong dominance of perturbative
effects if both photons are highly virtual. It turns out that for
$\gamma^*\gamma^*$ scattering this virtuality need not be too high. Already
for $Q_1^2, Q_2^2 \ge 5{\rm GeV}^2$ there is clear evidence for the
domination of the purely perturbative contribution. 

The models are discussed in detail in Section 2, results are presented in
Section 3 and final comments made in Section 4. At the end of Section 3 we 
present some simple fitting functions for the numerical results of our model.

We use the standard notation: $W = \sqrt{s}$ is the
$\gamma^{(*)}\gamma^{(*)}$ c.m. energy; $Q_i^2=- q_i^2$ are the photon
virtualities. If one photon is on shell and one off shell we denote by $Q^2$
the virtuality of off shell photon and $x=\frac{Q^2}{s+Q^2}$.

\section{The Model}

\subsection{The Pomeron Contribution}

For the colour singlet exchange we use an eikonal approach \cite{n} to high
energy scattering particularly suited to incorporate non-perturbative aspects
of QCD. The non-perturbative  behaviour of QCD is treated in the Model of
the Stochastic Vacuum \cite{d,ds} which approximates the IR part of QCD  by a
Gaussian stochastic process in the colour field strength. This model yields
linear confinent and can also be applied to high energy hadron-hadron
scattering, or more generally to quark-antiquark dipole-scattering \cite{dfk}. 
The model depends essentially on two typically non-perturbative parameters,
which specify the Gaussian process mentioned above: the strength of the gluon
correlator and $a$, the correlation length. These are related to the slope of 
the linear confining potential \cite{d,ds}. As it stands the model leads to 
cross sections which are constant with increasing energy. The parameters of 
the model were fitted to the iso-singlet exchange part of (anti-)proton-proton
scattering at $W = \sqrt{s}$= 20 GeV.  The phenomenologically observed 
increase with energy of hadronic total cross sections like $s^{(\alpha_P -1)}$
with $\alpha_P \approx 1.08$ \cite{dl2} can be incorporated in 
two ways: either one lets the radius of the hadrons increase with $s$ 
\cite{dfk} - \cite{BN98}, or one takes the model as a determination of the 
Regge residue and adds the Regge-like increase with energy by a factor 
$(s/s_0)^{(1-\alpha_P)}$ with $\surd s_0 = 20{\rm GeV}$. These two approaches 
give very similar results, and we adopt the latter in this paper as it is the 
more convenient in the present context.

Whereas hadron-hadron scattering and soft electroproduction processes (i.e. 
those with low photon virtuality $Q^2$ ) can be very well described in this 
way, it is well known the energy dependence for hard electroproduction
processes is much stronger than indicated by the soft non-perturbative pomeron.
As discussed in the Introduction the occurence of a second (hard) pomeron as
proposed in \cite{dl} can explain the data in a consistent way. This two 
pomeron approach was adapted to the MSV model in \cite{r} and very 
successfully tested for the electro- and photoproduction of vector mesons and,
more relevantly here, for the proton structure function over a wide range of
$x$ and $Q^2$. As in \cite{dl} it was found that the soft-pomeron contribution
to $F_2$, after an initial increase with increasing $Q^2$, has a broad maximum
in the region of 5 GeV$^2$ and then decreases as $Q^2$ increases further i.e.
it exhibits higher-twist like behaviour. In the context of the present model 
this is a consequence of the decreasing interaction strength with decreasing 
dipole size.

It is worth recalling the salient features of this version of the two-pomeron
model, to illustrate the distinction between the soft and the hard pomeron 
in dipole-dipole scattering. In \cite{r} it was assumed that all dipole
amplitudes in which both dipoles are larger than the correlation length
$a=0.35$ fm are dominated by the soft pomeron, and the energy dependence
therefore given by $(s/s_0)^{\alpha_{soft}-1}$ with $\surd s_0 = 20$ GeV and
$\alpha_{soft} = 1.08 + 0.25 t$. This ensures that the hard pomeron has
essentially no impact on purely hadronic scattering. If at least one of the
dipoles is smaller than $a=0.35$ fm then the trajectory is replaced by a fixed
pole $\alpha_{hard} =1.28$. This value was chosen as experimentally $F_2 \sim
s^{0.28}$ at $Q^2 = 20  {\rm GeV}^2$ and the fixed-pole approximation made
because of the lack of shrinkage in the $J/\psi$ photoproduction cross
section. It turned out that the model overestimated the non-perturbative
contribution of very small  dipoles so it was put to zero if either of the
dipoles is less than 0.16 fm. With only four parameters it was possible to
obtain a good description of data for the proton structure function and for
the electroproduction of vector mesons without noticeably affecting earlier
fits to hadron-hadron scattering. 

We apply this two-pomeron model without change to the evaluation of the
$\gamma^{(*)}\gamma^{(*)}$ cross sections. It should be noted that the 
simple factorisation formula $\sigma_{\gamma\gamma} = \sigma_{\gamma p}^2/
\sigma_{p p}$ is no longer applicable in the two-pomeron situation.

The considerations outlined briefly above lead to a model for the
scattering of quark-diquark dipoles on each other. In order to
relate it to $\gamma^{(*)}\gamma^{(*)}$ interactions we have to
introduce the photon wave function. In \cite{DGP98} it was shown by model
considerations that the lowest-order perturbative
expression for the quark-antiquark content of the photon,  with chiral
symmetry breaking and confinement being simulated by a $Q^2$-dependent quark
mass, works remarkably well.  The quark mass can be determined by comparing the
result for the vector-current correlator with the analytically
continued phenomenological expression  in the
Euclidean region. The resulting masses are:

\begin{eqnarray} \label{mass}
m_{u,d} & = & \left
\{ \begin{array}{r@{\quad:\quad}l} m_0 \,(1-Q^2/1.05) & Q^2 \le 1.05\\  0&
Q^2 \ge 1.05 \end{array} \right. \\ 
m_{s} & = & \left\{
\begin{array}{r@{\quad:\quad}l} 0.15 + 0.16\,(1-Q^2/1.6) & Q^2 \le 1.6\\  0.15
& Q^2 \ge 1.6 \end{array} \right. \nonumber\\ m_c & = & 1.3 \nonumber
\end{eqnarray}

The parameter $m_0$ for the $u,d$ quarks was found to be $m_0 = 0.21 \pm
0.015$ GeV.

\subsection{The Box Diagram}

For Compton scattering on hadrons it was shown in  
\cite{BGLL67a} - \cite{Sin67} that the Fubini-Dashen-Gell-Mann sum rules which 
relate the integral over the imaginary part of the Compton amplitude to the 
electromagnetic form factor $F(t)$ lead under very general assumptions to a 
fixed (i.e. t-independent) pole in the complex $J$-plane whose residue is 
proportional to $F(t)$. The residue is real and hence cannot contribute to 
the structure function. Such a 
fixed $J$-plane singularity also occurs in photon-photon scattering due to the
box diagramm. The large-$s$ behaviour of it is independent of the
momentum transfer and the virtuality and is of the order  $ i \log s$ which
corresponds to a fixed double pole.
For very large virtualities of the photons or high quark masses the QCD
corrections to the box diagram are under control \cite{HR79} and it
should not be modified by them in any essential way. It is thus natural
to add the contribution of the box diagram representing this
$J$-plane singularity to the valence quark
contributions corresponding to the Reggeon exchange without conceptual
difficulties of double counting. The singularity contributes to the
imaginary part of the residue and we have a contribution to the
gamma-gamma cross section 
$\sigma_{\gamma^*\gamma^*}^{box} = \mbox{const }\times \log(s)/s$.

We give here the full cross section for
$\gamma^{(*)}\gamma^{(*)}$ scattering for a colour triplet of quarks
with mass m and charge $ e_f= \hat e_f \cdot e $:
\begin{eqnarray} \label{bd}
&& \hspace{-7mm}\sigma(W^2,Q_1^2,Q_2^2) = -\frac{3\pi}{2}  {\hat e}_f^4  \alpha^2 \frac{|\vec p\,|}
{|\vec q\,| W^2}\\
&&\hspace{-5mm}\times\left( 8 + {\frac{4\,
        \left( 2\,{m^2} - {Q_2^2} \right)
          \,\left( 2\,{m^2} - {Q_1^2}
           \right) }{{|\vec p|}\,
        {|\vec q|}\,
        \left( {Q_2^2} - 
          4\,{|\vec p|}\,{|\vec q|} + 
          {Q_1^2} + W^2 \right) }} - 
    {\frac{4\,\left( 2\,{m^2} - 
          {Q_2^2} \right) \,
        \left( 2\,{m^2} - {Q_1^2} \right) 
        }{{|\vec p|}\,{|\vec q|}\,
        \left( {Q_2^2} + 
          4\,{|\vec p|}\,{|\vec q|} + 
          {Q_1^2} + W^2 \right) }} \right. \nonumber\\
&&\hspace{-5mm} \left.    +{\frac{2\,\left( -8\,{m^4} + 
          {Q_2^4} + 
          2\,{Q_2^2}\,{Q_1^2} + 
          {Q_1^4} + {W^4} + 
          4\,{m^2}\,\left( {Q_2^2} + 
             {Q_1^2} + W^2 \right)  \right)}{{|\vec p|}\,{|\vec q|}\,
        \left( {Q_2^2} + 
          {Q_1^2} + W^2 \right) }} 
         \, \right. \nonumber \\ 
&& \left. \times\log ({\frac{{Q_2^2} - 
             4\,{|\vec p|}\,{|\vec q|} + 
             {Q_1^2} + W^2}{{Q_2^2
              } + 4\,{|\vec p|}\,
              {|\vec q|} + {Q_1^2} + W^2
             }})\right)\nonumber
\end{eqnarray}

with 

$${{{|\vec p|}}= 
    {{\sqrt{-{m^2} + {\frac{W^2}{4}}}}}},
~~~~{{{|\vec q|}}=
    {{\frac{{\sqrt{{Q_2^4} - 
            2\,{Q_2^2}\,{Q_1^2} + 
            {Q_1^4} + 
            2\,{Q_2^2}\,W^2 + 
            2\,{Q_1^2}\,W^2 + {W^4}}}}{2\,
         {W}}}}}.$$ 

If $m_q^2 \ll Q_1^2 \ll W^2$ and $Q_2^2 \ll m_q^2$ we have:
$$\sigma_{\gamma^*\gamma^*} = 4 \pi^2 \alpha^2 F_2^\gamma = 12 \pi \alpha^2 
\left(\sum_f {\hat e_f}^2\right)^2
{{1}\over{W^2}} \left(\log\frac{W^2}{m_q^2}-1\right).$$

For the case: $0 \ll Q_1^2, Q_2^2 \ll W^2$ we have:
$$\sigma_{\gamma^*\gamma^*} = 4 \pi^2 \alpha^2 F_2^\gamma = 12 \pi \alpha^2 
\left(\sum_f {\hat e_f}^2\right)^2
{{1}\over{W^2}} \left(\log\frac{W^4}{Q_1^2 \,Q_2^2}-1\right).$$

If at least one of the photon virtualities is smaller than the internal 
quark mass the box diagram receives important contributions from the IR 
region and thus depends crucially on the quark mass. In our approach it is 
natural to use therefore for small $Q^2$ the same $Q^2$-dependent quark mass 
as in the photon wave function (see equation \ref{mass}).

\subsection{{\bbf The Reggeon}}

In many respects the contribution from the coupling of the
reggeon to the hadronic content of the photon ($\rho$, $\omega$, $\phi$ etc.)
is the least well-defined. Even with one  photon on-shell, i.e. for the valence
quark contribution $F_{2,had}^{\gamma}$  to the hadronic structure function of
the real photon, there are considerable  ambiguities \cite{grv1,afg,sas}. In
naive Vector Meson Dominance (VMD) this is  given by

\be
{{1}\over{\alpha}}F_{2,val}^\gamma(x,Q^2) = F_{val}^\pi(x,Q^2)
\sum_V{{4\pi}\over{f_V^2}},
\ee

where the sum is usually over $\rho$, $\omega$ and $\phi$. The additional 
assumption has been made that the vector meson structure functions can
all be represented by the valence structure function of the pion 
$F_{val}^\pi(x,Q^2)$. This in itself is quite an extreme statement, as 
there is no obvious reason why the structure function of the short-lived 
vector mesons should be the same as those of the long-lived pion. 
Additionally it is not clear whether one should take the simple incoherent 
sum or allow for coherence effects. Finally, higher-mass vector mesons must 
also make some contribution, but this is almost certainly small for the real 
photon as compared to the uncertainties in any estimate of the hadronic 
component.

To add to these uncertainties, the pion structure function is only known
experimentally for $x > 0.2$. To obtain the structure function in the
kinematical domain of interest here, it is necessary to use the DGLAP 
evolution equations to fit the data and to extrapolate \cite{aur,grv2}.  
This was the approach used by \cite{grv1,afg} in fitting $F_2^\gamma$, 
although with somewhat different assumptions about the effective strength
of the contribution. In contrast, in \cite{sas} the shape of the
hadronic contribution was left free to be determined by the data, but 
the normalisation was fixed. 

In our previous work \cite{ddr} we used the DGLAP evolved pion structure 
function of \cite{grv2}, and retained only the $\rho$, $\omega$ and $\phi$
in the sum of eqn.(1). At small $x$ ($x \leq 0.1$ ) and small $Q^2$ 
($Q^2 \leq 25 {\rm GeV}^2$ ) this
can be well  parametrised by

\be
F_{2,val} = C\bigl({{Q^2}\over{Q^2+a}}\bigr)^{1-\eta}x^\eta
\ee

with $a = 0.3$ GeV$^2$, $C = 0.38$ and $\eta = 0.45$. Thus for the valence
quark contribution to the $\gamma^* \gamma$ cross section we get

\be
\sigma_{\gamma^* \gamma}(s,Q^2) = 4\pi^2\alpha^2{{C}\over{s}}
\bigl({{s}\over{Q^2+a}}\bigr)^{1-\eta} = 
{{312}\over{s}}\bigl({{s}\over{Q^2+0.3}}\bigr)^{0.55} {\rm nb.}
\ee

This simple formula holds to better than $10\%$ over the $(x,Q^2)$ range 
relevant for LEP. That is the error is much less than the other uncertainties 
in estimating this term.

Extrapolating eqn.(3) to $Q^2 = 0$ does not satisfy simple factorisation. The 
total $\gamma p$ and $p p$ ($p\bar p$) cross sections can be described by 
three terms corresponding to soft pomeron exchange, $C = +1$ reggeon exchange 
and $C = -1$ reggeon exchange, with universal powers of $s$ for each. In the
absence of cuts (and the universality implies that these should be small)
each term should factorise independently. That is

\be
\sigma^i_{\gamma \gamma} = {{(\sigma^i_{\gamma p})^2}\over{\sigma^i_{p p}}}
\ee

where $i$ corresponds to any one of the soft pomeron, $C = +1$ reggeon or 
$C = -1$ reggeon contributions. The value of $\eta$ used in eqns.(2) and (3) 
corresponds to the fit to total cross sections of \cite{dl2}, from which one 
finds

\be
\sigma^{C=+1}_{\gamma \gamma} = 216 s^{-\eta} {\rm nb}
\ee

which is about ${{1}\over{3}}$ of eqn.(3) in the $Q^2 \ra 0$ limit. If one 
applies factorisation to the latest PDG fit \cite{pdg} to total cross sections
then the result is approximately midway between these two extremes although 
with a somewhat different energy dependence. Thus it seems reasonable, given 
all the uncertainties, to take eqns.(3) and (5) as giving upper and lower 
limits respectively to the reggeon exchange contribution to the total hadronic 
$\gamma \gamma$ cross section.

The simplest approach to extending eqn.(3) to the case when both photons 
are off-shell is to assume that as it is a reggeon contribution it should 
factorise:

\be \label{reggeon}
\sigma_{\gamma^* \gamma^*}(s,Q_1^2,Q_2^2) = 
4\pi^2\alpha^2{{C}\over{a}}\bigl({{a}\over{Q_1^2+a}}\bigr)^{1-\eta}
\bigl({{a}\over{Q_2^2+a}}\bigr)^{1-\eta}
\bigl({{s}\over{a}})^{-\eta} {\rm nb.}
\ee

with $C$ and $a$ having the same values as before.

\section{Results}

\begin{figure}[p]
\leavevmode
\centering
\begin{minipage}{7.5cm}
\epsfxsize7.5cm
\epsffile{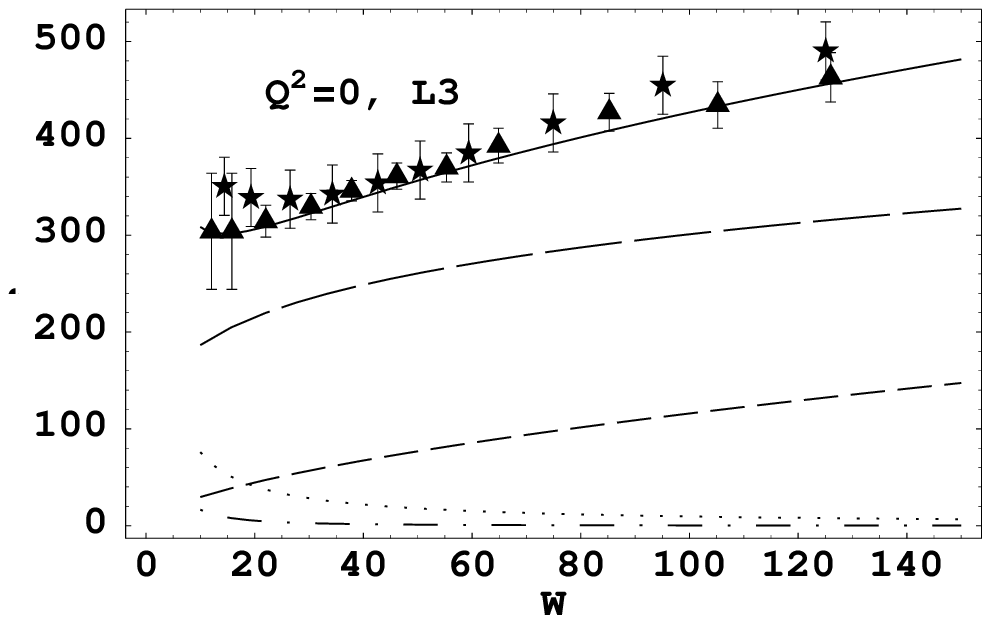}
\centering  
\end{minipage}
\hfill
\begin{minipage}{7.5cm}
\epsfxsize7.5cm
\epsffile{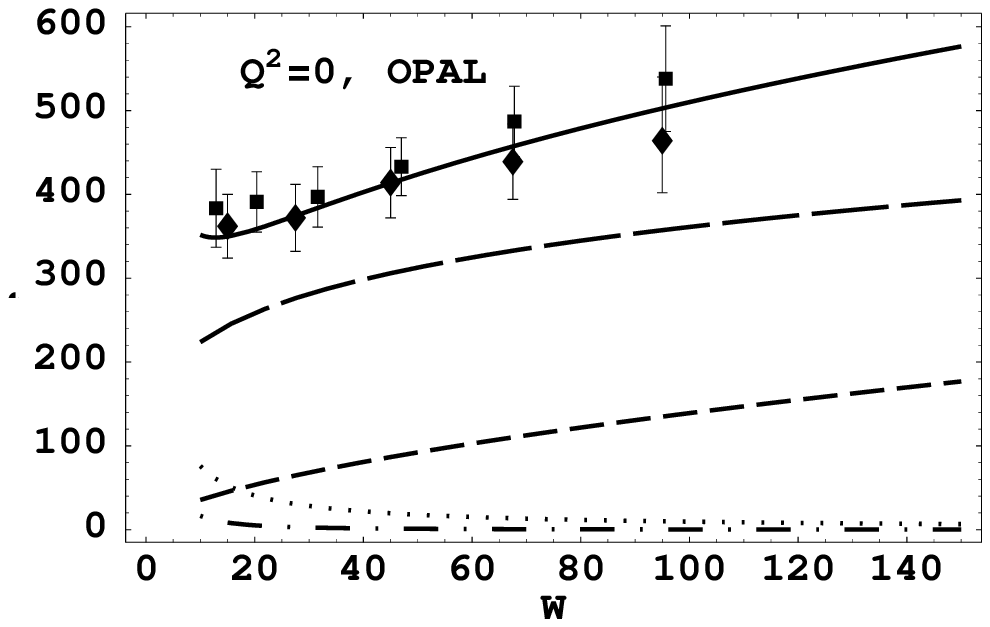}
\centering 
\end{minipage}
\begin{minipage}{7.5cm}
\epsfxsize7.5cm
\epsffile{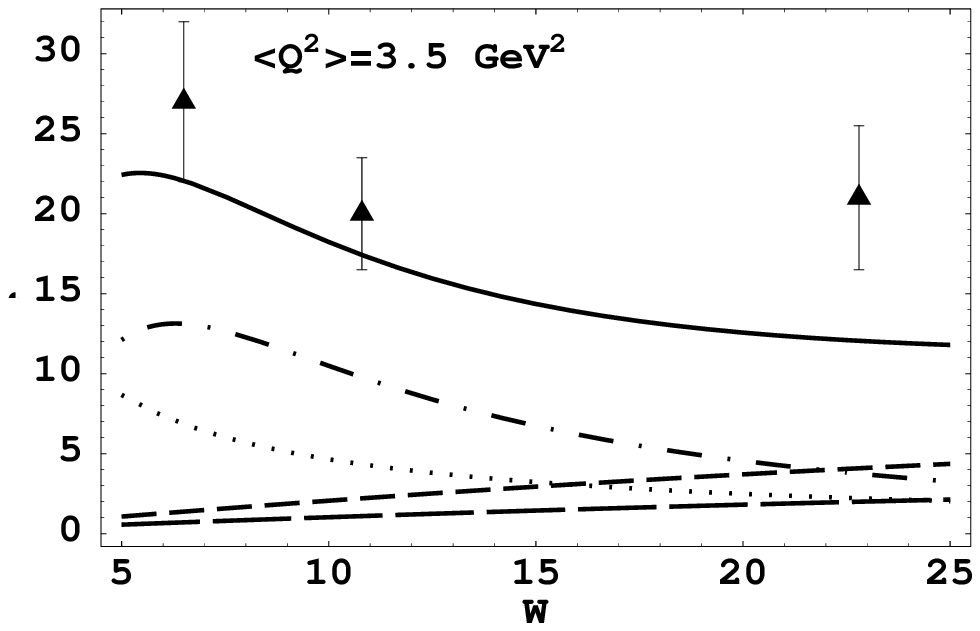}
\centering  
\end{minipage}
\hfill
\begin{minipage}{7.5cm}
\epsfxsize7.5cm
\epsffile{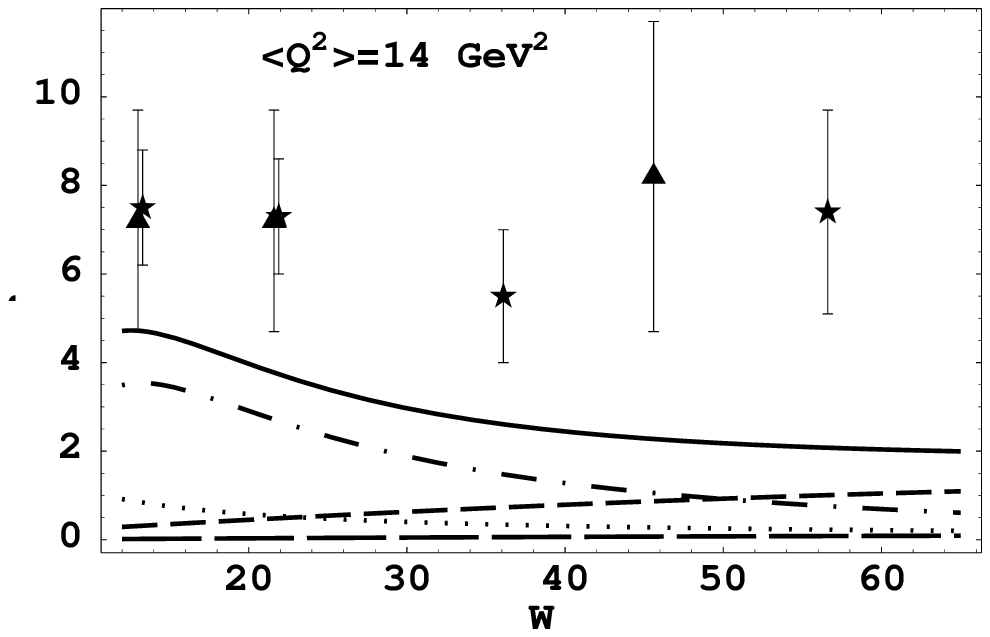}
\centering 
\end{minipage}
\caption{Cross sections in nb for $\gamma^{(*)}\gamma^{(*)}$ scattering
for virtualities $Q^2$ = 0, 3.5 and 14 GeV$^2$ respectively compared with
OPAL and L3 data.
L3 \protect \cite{L3a,L3b}, Triangles; L3 \protect \cite{L399d} and
private communication, Stars:
OPAL \protect \cite{OPALa}, Boxes ; OPAL \protect \cite{OPA99a}, Diamonds. 
The solid curve is our model . It consists of the
following contributions: soft pomeron: long dashes; hard pomeron: short dashes;
fixed pole (box): dot-dashes; reggeon: dots.
For the L3 data a $Q^2$-dependent quark mass with $m_0 =
0.21$ GeV was used, for the OPAL data $m_0 = 0.20 $ GeV, see equation \protect
(\ref{mass}).} \end{figure}
%%%%%%%%%%%%%%%%%%%%%%%%%%%%

The pomeron contribution to $\sigma_{\gamma\gamma}$is rather
sensitive to the effective light-quark mass $m_q$ entering the photon
wave function, varying as $\sim 1/m_q^4$. This is illustrated in Figs.(1a) 
and (1b) which show separately the L3~\cite{L3a,L399d} and
OPAL~\cite{OPALa,OPA99a} data  and the pomeron model with $m_q = 210$ MeV and
200 MeV respectively, together  with the other contributions to the total
cross section. These values of  $m_q$ are slightly lower than the 220 MeV used
in the earliest calculations,  but have no observable effect on the purely
hadronic predictions of the model  and actually serve to improve slightly the
description of high-energy  photon-proton reactions~\cite{DGP98,kdp98}. They
are also within the expected  range of $210 \pm 15$ MeV, as determined from
the two point vector function \cite{DGP98}. The sensitivity to $m_q$
disappears once $Q^2 \gg m_0^2$. 

It is clear that the hard part of the pomeron contribution becomes 
increasingly important with increasing energy, reaching more than 20$\%$ of 
the cross section at a c.m. energy of 130 GeV. This relatively strong fraction
of the hard part is a consequence of the pointlike coupling of the photons to 
the quarks and the resulting singularity at zero distance of the photon wave 
function. In the corresonding picture for proton-antiproton scattering the 
hard part is only about $1\%$ of the cross section at $W=130$ GeV.   

The predictions of the same model for the $\gamma^*\gamma^*$ cross sections 
at $Q^2$ = 3.5 GeV$^2$ and 14 GeV$^2$ are compared with the recent L3 data 
\cite{L3b} in Table \ref{tab1} and shown in Fig.2. The predictions at $Q^2$=
3.5  GeV$^2$ are slightly below and at 14 GeV$^2$ distinctly below the data, 
especially at the higher values of $W$. This and the many successful tests 
of the soft pomeron part within the model make it very likely that the 
discrepancy is due to a wrong $Q^2$ dependence of the hard part. Therefore 
in Fig.3 we plot the difference between experiment and the sum of soft 
pomeron and non-diffractive terms. This difference represents the hard part 
of the reaction. It is interesting that at all virtualities the data can be 
fitted well with a power behaviour $W^{2\epsilon} = s^{\epsilon}$ with
$\epsilon \sim 0.3$, fully consistent with a hard second pomeron. Of course
the error on $\epsilon$ is large $\sim 0.1$. The comparatively small 
contribution to the cross section from the soft pomeron is directly 
attributable to its decreasing interaction strength with decreasing dipole 
size (the higher-twist behaviour) found in deep inelastic scattering 
\cite{dl,r}.

\begin{table}
\begin{center}
\begin{tabular}{|c|c|c|c|c|c|c|c|}
\hline
$W$&L3& F.P. & R.P. &{S.P.} & {H. Ex.} & Error & 
{H. Mod.}\\ \hline
\multicolumn{8}{|l|}{$Q^2=3.5$} \\
6.5& 27& 13.1& 6.8& 0.7& 6.3& 5& 1.4\\ 
  10.8& 20& 9.8& 4.3& 1.1& 4.8& 3.5& 2.2\\ 
  22.8& 21& 3.8& 2.2& 2.& 13.0& 4.5& 4.1\\
\hline
\multicolumn{8}{|l|}{$Q^2=14$} \\
13& 7.2& 3.5& 0.9& 0.02& 2.8& 2.5& 0.32\\ 
21.6& 7.2& 2.7& 0.5& 0.04& 3.9& 2.5& 0.48\\ 
45.6& 8.2& 1.1& 0.3& 0.07& 6.8& 3.5& 0.87\\
\hline
\multicolumn{8}{|l|}{$Q^2=14.5$}\\
13.3&7.5& 3.4&0.8&0.02&2.5&1.3*&0.30\\
21.9&7.3&2.6&0.5&0.03&4.17&1.3*&0.48\\
36.1&5.5&1.5&0.3&0.05&3.65&1.5*&0.68\\
56.6&7.4&0.8&0.2&0.08&6.3&2.3*&0.99\\
\hline
\end{tabular}
\end{center}
\caption{$\gamma^*\gamma^*$ total cross section in nb.
$W$: $\gamma^{(*)}\gamma^{(*)}$ c.m. energy [GeV];
L3: experimental results from L3 \cite{L3a,L3b};
F.P.: fixed pole (box-diagram);
R.P.: reggeon-contribution (valence term);
S.P.: soft pomeron (non-perturbative contribution);
H.Ex: hard `experimental' contribution:
Error: experimental error;
H.Mod: hard contribution extrapolated from  the model adapted
to the proton structure function \protect \cite{r}.\label{tab1}}
\end{table}

\begin{figure}[p]
\leavevmode
\centering
\begin{minipage}{7.5cm}
\epsfxsize7.5cm
\epsffile{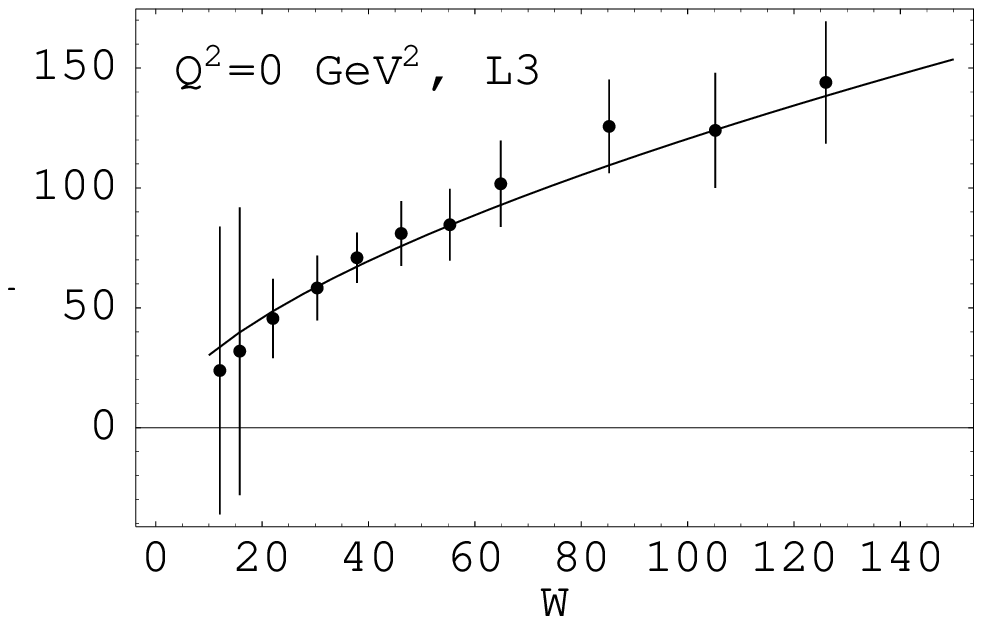}
\centering  
\end{minipage}
\hfill
\begin{minipage}{7.5cm}
\epsfxsize7.5cm
\epsffile{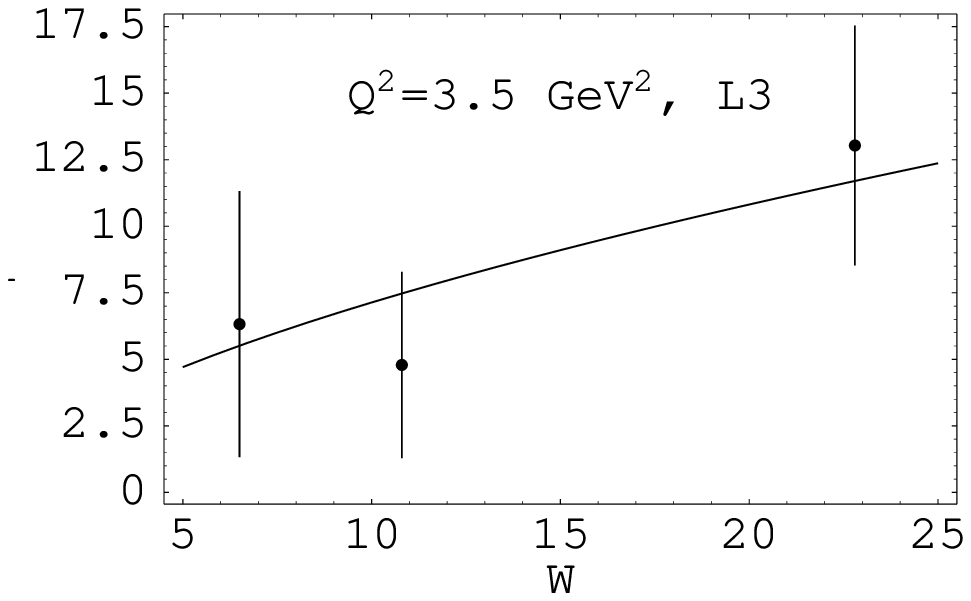}
\centering 
\end{minipage}
\begin{minipage}{7.5cm}
\epsfxsize7.5cm
\epsffile{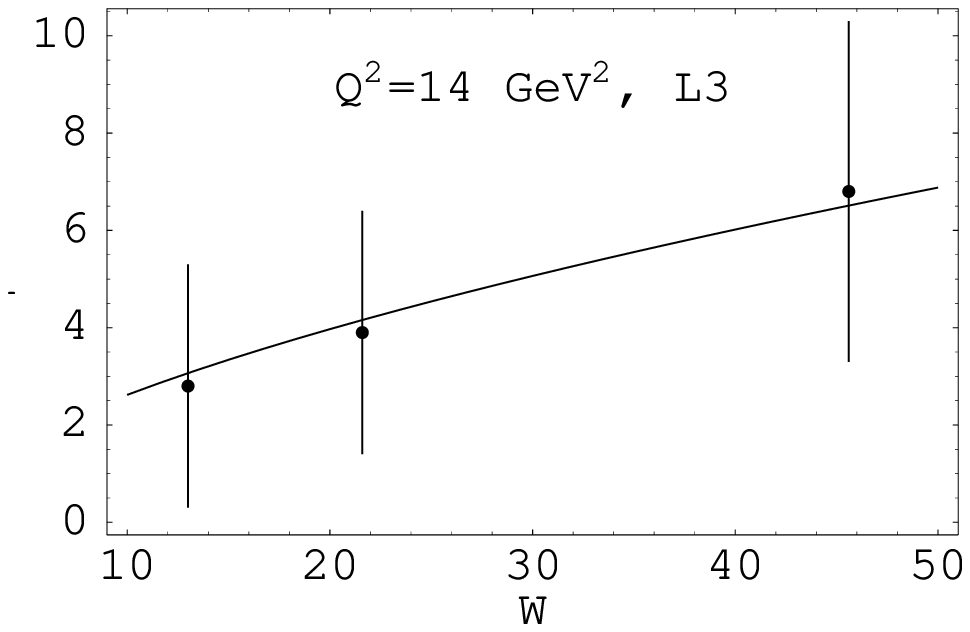}
\centering  
\end{minipage}
\begin{minipage}{7.5cm}
\epsfxsize7.5cm
\epsffile{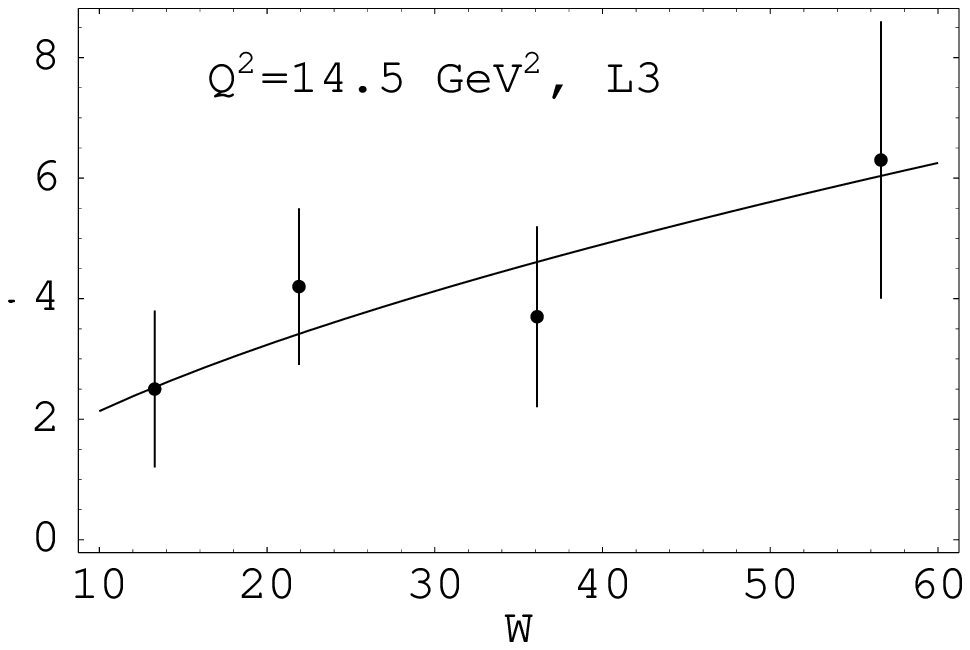}
\centering  
\end{minipage}
\caption{ L3 data \protect \cite{L3a,L3b} with the soft pomeron, reggeon,
and fixed pole (box) contributions subtracted. The solid line is a fit of the
form $A\times (W/GeV)^{0.6}$ with $A= 7.6,1.8,0.66,0.54 $ nb for $Q^2$ 
= 0, 3.5, 14, and 14.5  GeV$^2$ respectively.}  
\end{figure} 

\begin{table}
\centering
\begin{tabular}{|l| l| l| l|}
\hline
& {Non-pert.}& {Pert.} & naive VDM \\
\hline
2 dipoles  & 
{$R_1^2 \cdot R_2^2$} &{$R_1 \cdot R_2$ }&\\

small:&{$\frac{1}{Q_1^2\cdot Q_2^2} $}& 
 {$\frac{1}{Q_1\cdot Q_2}$} &
$\frac{1}{Q_1^4\cdot Q_2^4}$\\ 
&&&\\
\hline

1 dipole  & 
{$R_i^2$} & {$ R_i^2$} & \\
small: &
$\frac{1}{Q_i^2}$&
  {$\frac{1}{Q_i^2}$}&
$\frac{1}{Q_i^4}$\\ 
&&&\\
\hline
\end{tabular}
\centering
\caption{Behaviour of the different contributions leading in W 
to the $\gamma^*\gamma^*$ cross section (up to
logarithmic terms).\label{tab2}} 
\end{table}

We can easily see why the model with parameters taken from the nucleon
structure functions can fail when applied to the $\gamma^*\gamma^*$ cross 
section. In Table \ref{tab2} we give the power dependence on the dipole size
$R$, and  correspondingly $Q^2$, of different terms: perturbative two gluon
exchange:  the genuine non-perturbative contribution: and the contribution of
(naive)  vector meson dominance. We see that for one dipole large (e.g. the
nucleon  in the structure functions) the $Q^2$ behaviour is the same for the
total pomeron contribution in the model and the perturbative two-gluon
exchange. Therefore a distinction between the two is difficult. However this
is not the  case when the two dipoles become small. Here the perturbative
two-gluon  exchange falls off much slower with decreasing dipole size
(increasing $Q^2$) than does the pomeron contribution in the model. It is
therefore very plausible that the residue of the hard pomeron has a $Q^2$
dependence more akin to that of the perturbative contribution rather than to
the nonperturbative term (as  implied in the present model~\cite{r}).

If this explanation is correct then the present model should still give a 
good description of the structure function of the real photon, $F_2^{\gamma}$, 
as the real photon is dominated by a large dipole size. It is similar to, but 
not exactly the same as, the nucleon structure function but is not precisely 
analogous as the real photon has a small-dipole component not present in the 
nucleon. The predictions of the model are compared with $F_2^{\gamma}$ data in
Fig.4. They extend to larger $x$ than in our previous work
\cite{ddr} because of the inclusion of the box diagram which we omitted 
previously : $x \leq {\rm min}(0.2,Q^2/(25+Q^2)$, i.e $W \geq 5$ GeV. 
The agreement with experiment is indeed very satisfactory stressing again the
reliability of the model if at least one of the dipoles is large. Figure 5
shows the smooth extrapolation from the purely perturbative domain ($Q^2 =0$)
to the domain of DIS. It can be seen from the prediction for $W=50$ GeV 
that at that energy the hard pomeron is dominant even at moderate
virtualities. This  stresses the relevance of $F_2^{\gamma}$ as data can 
be taken  at smaller $x$ (higher $W$) and larger $Q^2$ than for the 
$\gamma^*\gamma^*$  cross section, so it remains a sensitive probe of the 
hard pomeron.
\begin{figure}[p]
\leavevmode
\centering
\begin{minipage}{7.5cm}
\epsfxsize7.5cm
\epsffile{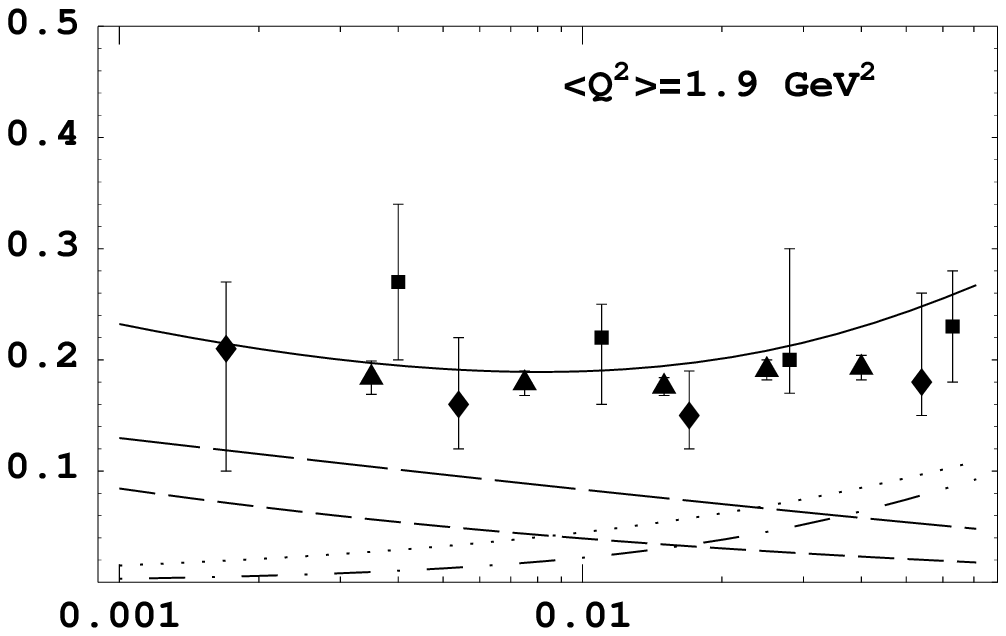}
\centering  
\end{minipage}
\hfill
\begin{minipage}{7.5cm}
\epsfxsize7.5cm
\epsffile{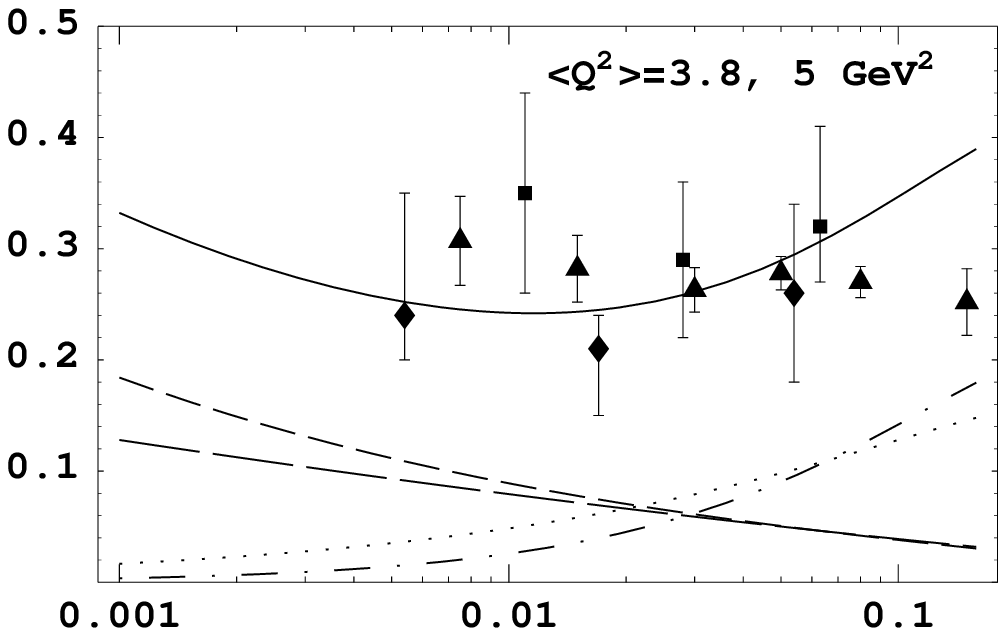}
\centering 
\end{minipage}
\begin{minipage}{7.5cm}
\epsfxsize7.5cm
\epsffile{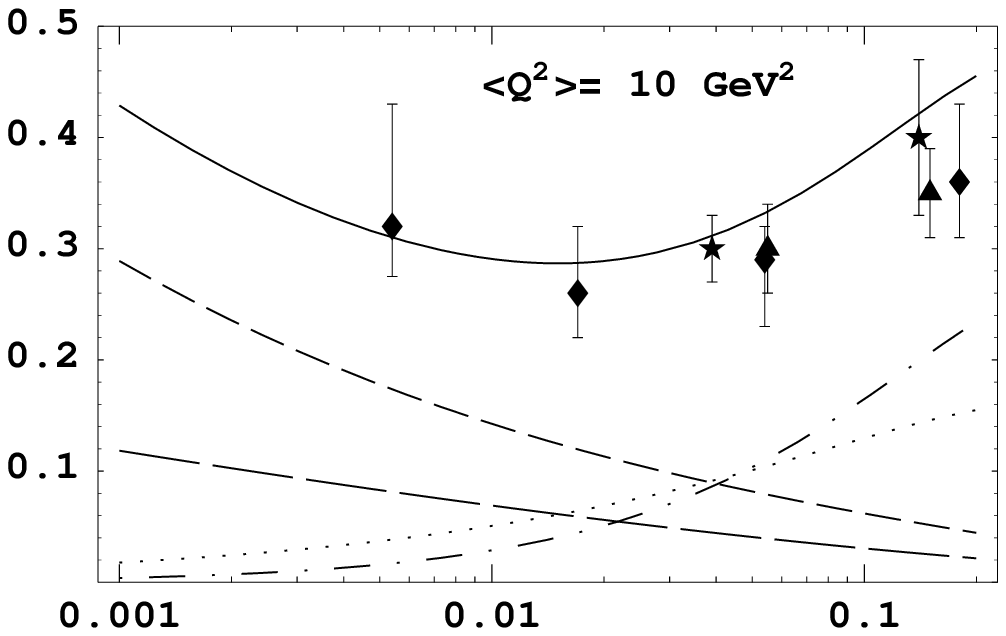}
\centering  
\end{minipage}
\hfill
\begin{minipage}{7.5cm}
\epsfxsize7.5cm
\epsffile{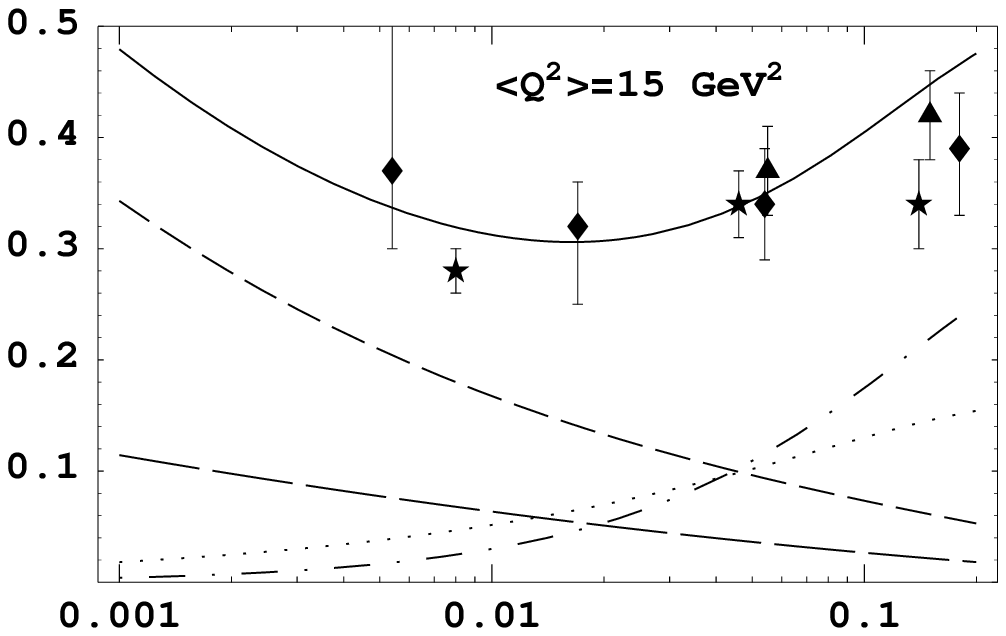}
\centering 
\end{minipage}
\begin{minipage}{7.5cm}
\epsfxsize7.5cm
\epsffile{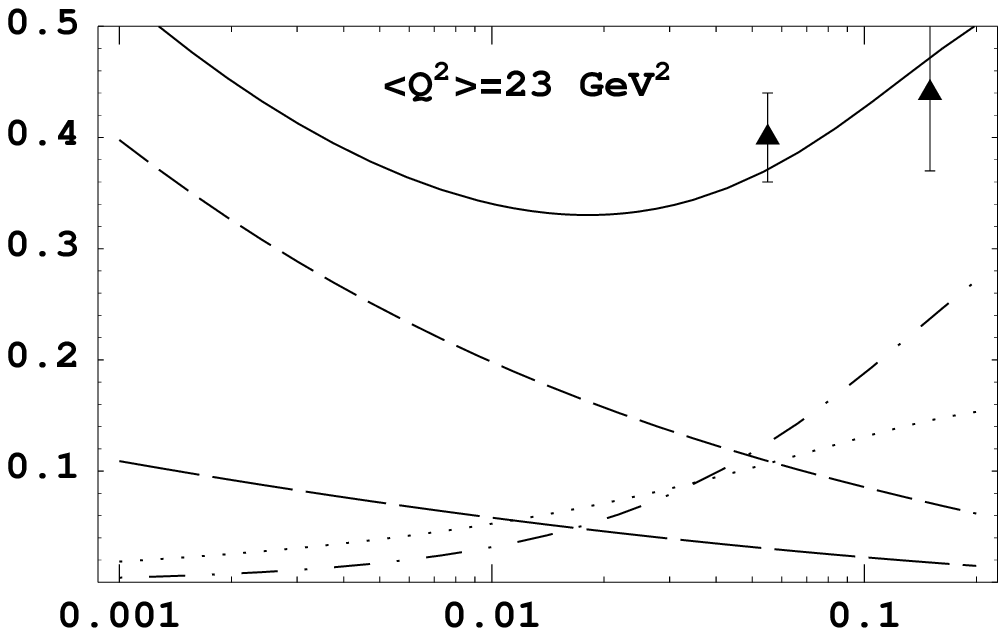}
\centering  
\end{minipage}
\caption{ The photon structure function $F_2^\gamma(x,Q^2)$ as function of
$x=\frac{Q^2}{W^2+Q^2}$. 
The data are: L3 \protect \cite{L398d}, Triangles: OPAL \protect
\cite{OPA97a}, Boxes;  OPAL \protect \cite{OPA99b},
Diamonds:
ALEPH \protect \cite{ALE99a,ALE99b}, Stars. 
The solid
curve is our model without adjusted parameters. It consists of the following
contributions: soft pomeron, long dashes; hard pomeron, short dashes; fixed 
pole, dot-dashes; reggeon, dots. 
The OPAL data in the $Q^2=5$ GeV$^2$ figure are at
$\langle Q^2 \rangle = 3.8 {\rm GeV}^2$;
the OPAL data in the $Q^2=15$ GeV$^2$ figure are at $\langle Q^2 \rangle = 17.6
{\rm GeV}^2$; and
the ALEPH  data in the $Q^2=15$ GeV$^2$ figure are at $\langle
Q^2 \rangle = 14 {\rm GeV}^2$}
\end{figure}

\begin{figure}[p]
\leavevmode
\centering
\begin{minipage}{7.5cm}
\epsfxsize7.5cm
\epsffile{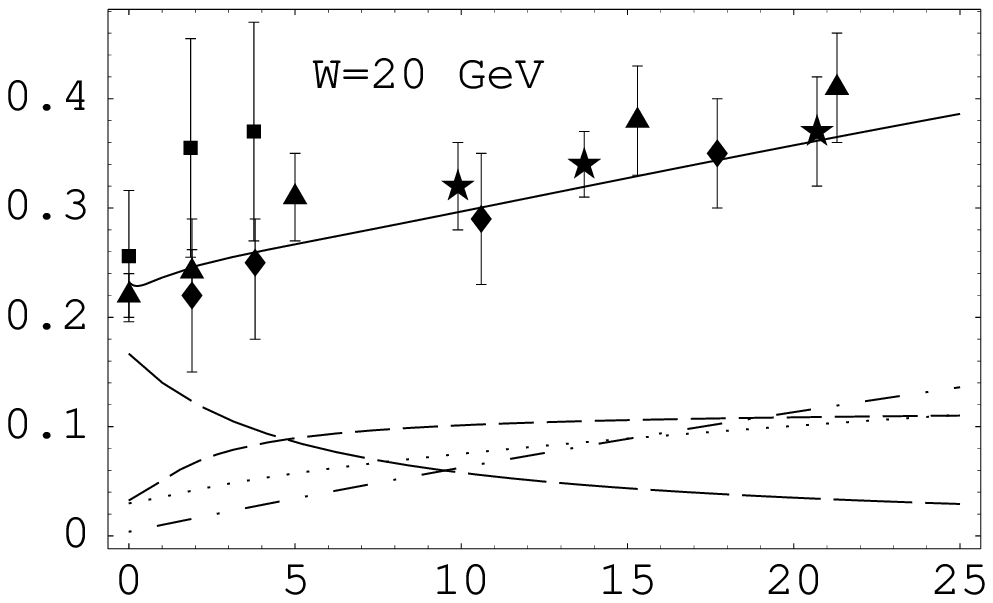}
\centering
\end{minipage}
\centering
\begin{minipage}{7.5cm}
\epsfxsize7.5cm
\epsffile{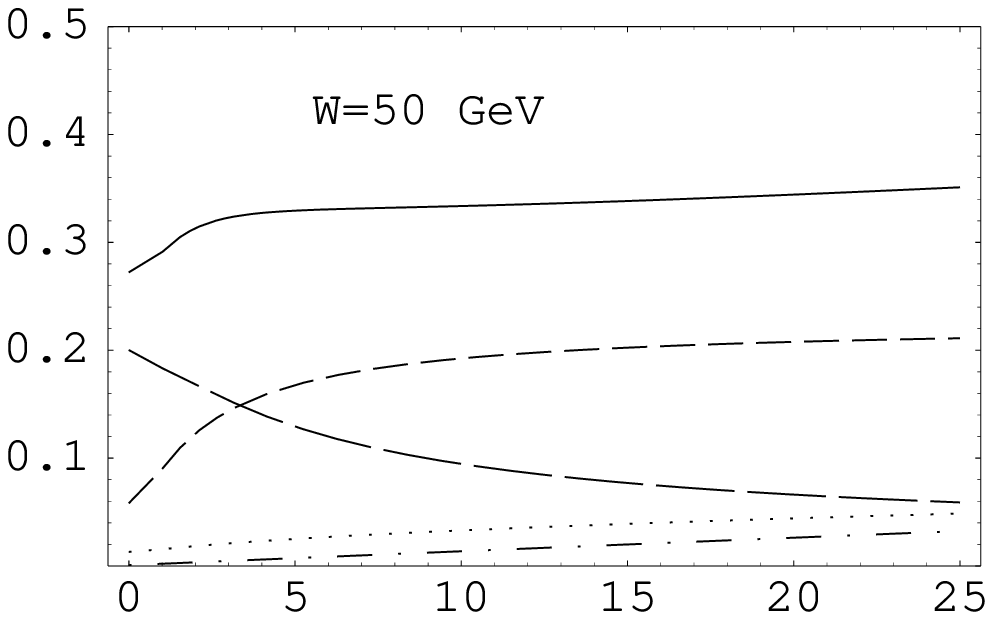}
\centering
\end{minipage}
\caption{The modified photon structure function $\tilde
F_2^\gamma=\frac{Q^2+0.6 {\rm
GeV}^2}{Q^2}*F_2^\gamma(x=\frac{Q^2}{W^2+Q^2},Q^2)$ as function of
$Q^2$ for $W \approx 20$ GeV and 50 GeV.  The data are: L3 \protect
\cite{L398d}, Triangles: OPAL \protect \cite{OPA97a}, Boxes; OPAL
\protect \cite{OPA99b},  Diamonds: ALEPH \protect
\cite{ALE99a,ALE99b}, Stars. The solid curve is our model without adjusted
parameters. It consists of the following contributions: soft pomeron, long
dashes; hard pomeron, short dashes; fixed pole, dot-dashes; reggeon, dots.}
\end{figure}

\begin{figure}[p]
\leavevmode
\centering
\begin{minipage}{7.5cm}
\epsfxsize7.5cm
\epsffile{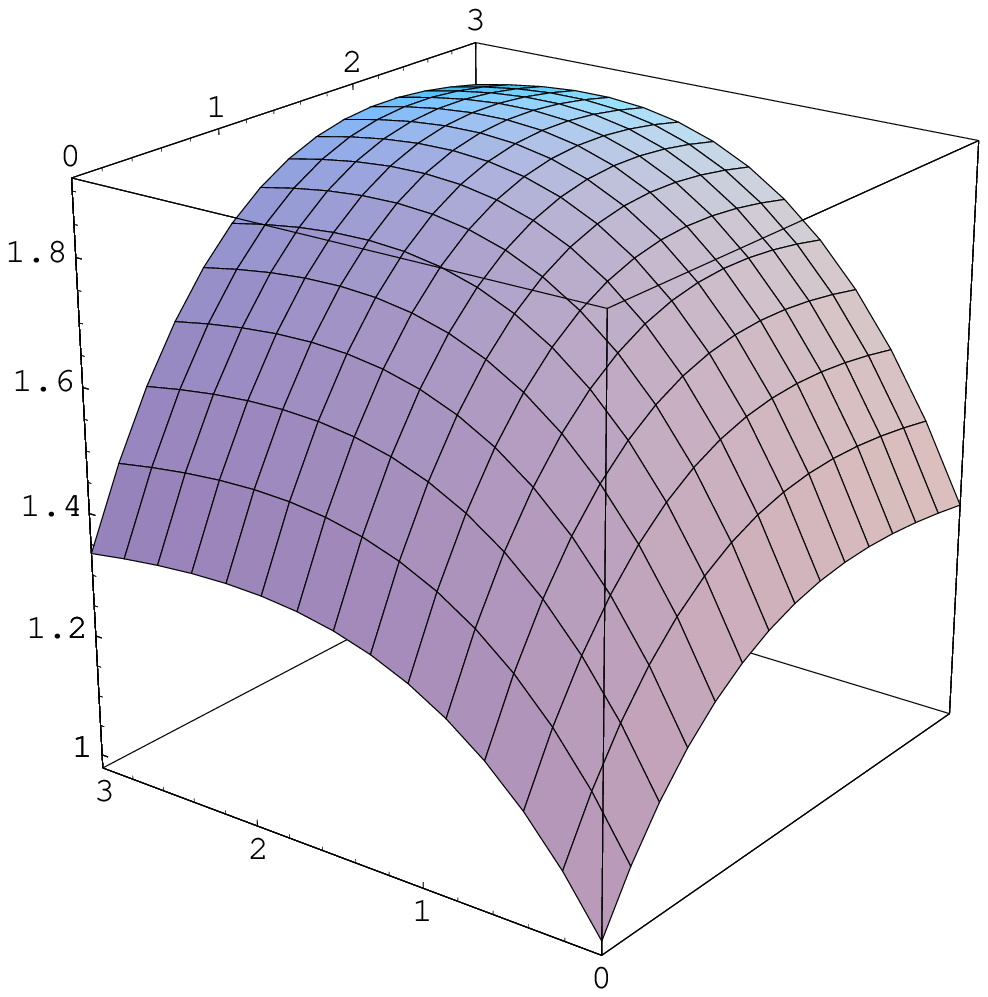}
\centering  $W=90$ GeV
\end{minipage}
\hfill
\begin{minipage}{7.5cm}
\epsfxsize7.5cm
\epsffile{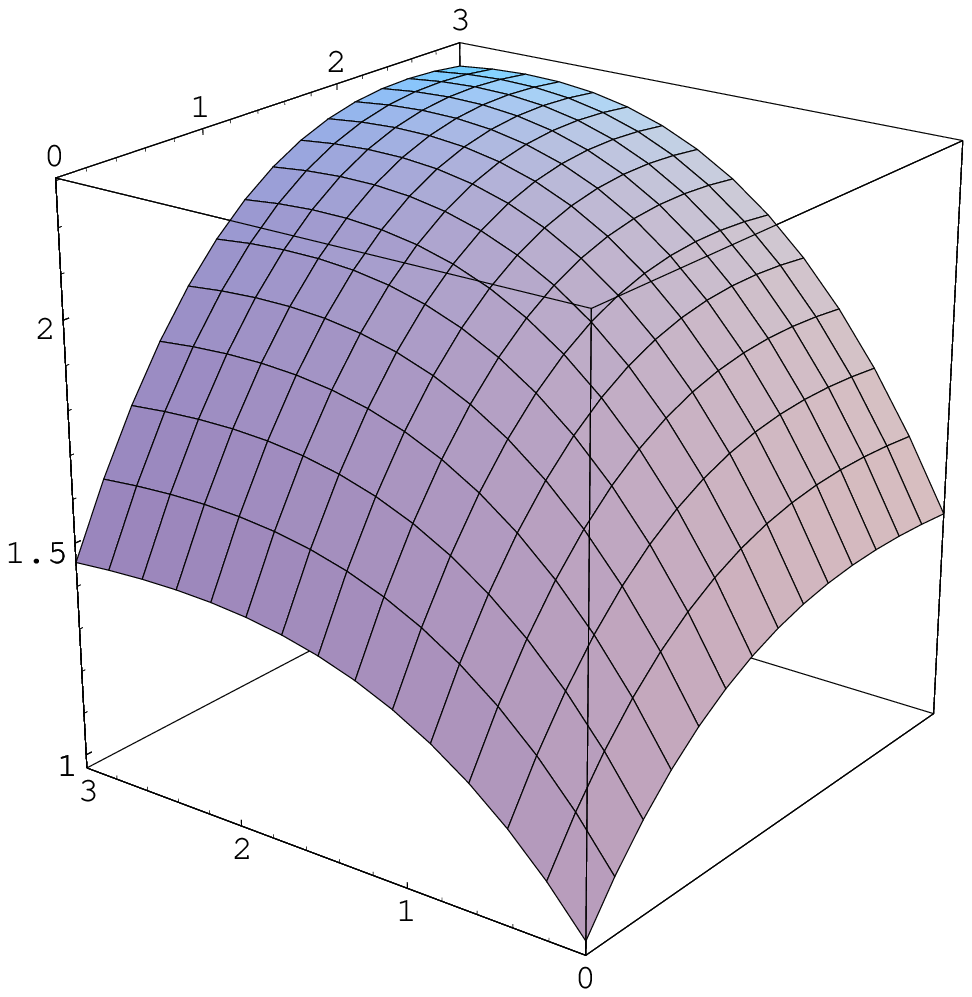}
\centering  $W=245$ GeV
\end{minipage}
\caption{Ratio of the model cross
  section to the naive VMD cross section as a function of $Q_i^2$ for
  $0 \le Q_i^2 \le 3$ GeV$^2$, normalized to one at $Q^2 = 0$. The two plots 
  are for the two centre of mass energies of the $\gamma^* \gamma^*$
  system $W=90$ GeV and $W=245$ GeV.}
\label{derivationVMD}
\end{figure}

Table \ref{tab2} demonstrates that the model shows a significant deviation from
simple  VMD at large $Q^2$, and this is still valid at small $Q^2$. A
comparison of the model and naive VMD is made in Fig.\ref{derivationVMD} which
shows the  ratio of the model cross section to the VMD cross section as a
function of  $Q^2$, normalised to one at $Q^2 = 0$. The two plots are for the
two centre  of mass energies of the $\gamma^*\gamma^*$ system $W=90$ GeV and
$W=245$ GeV.

This deviation from naive VMD is of particular importance in the
present evaluation of the real photon cross section,
$\sigma_{\gamma\gamma}$, as the data are untagged and cover a finite
range of $Q^2$. The normal procedure is to assume VMD to extract
$\sigma_{\gamma\gamma}$, but our results indicate that this is not
reliable. We fitted our results for the total
$\sigma_{\gamma^*\gamma^*}$ cross section in the kinematical range
$90<W<250$ GeV and $0<Q_i^2<3 \mbox{ GeV}^2$ with an ansatz which
shows explicitly the deviation from the naive VMD behaviour for the
$Q^2$ dependence and for fixed virtualities it is a simple power fit
for the $W$-dependence:
\beqa
\sigma_{\gamma^* \gamma^*}&=&\left(a+\left(b+c*Q_1^2\right)\right)*\exp\left(-d*Q_1^2\right)* \left(a+\left(b+c*Q_2^2\right)\right)*\exp\left(-d*Q_2^2\right)\nn\\
&&*1/\left(Q_1^2+m_\rho^2\right) * 1/\left(Q_2^2+m_\rho^2\right)\nn\\
&&*\left(W/20\mbox{ GeV}\right)^{\left(e+f*Q_1^2*\exp\left(-g*Q_1^2\right)\right) * \left(e+f*Q_2^2*\exp\left(-g*Q_2^2\right)\right)}\;.
\eeqa
Our fit for the seven parameters $a$-$g$ results in ($Q_i^2$ in $\mbox{GeV}^2$, $W$ and $m_\rho$ in GeV, $\sigma_{\gamma^* \gamma^*}$ in nano-barns)
\be
a=9.10,
b=0.398, 
c=3.380,
d=0.541,
e=0.554,
f=0.115,
g=0.276\;.
\ee

For convenience we give also fitting formulae for the case 
 of one real and one real or virtual
photon as obtained in the model applicable in the range 10 GeV $\leq W
\leq$ 150 GeV and $0 \leq Q^2 \leq 25 $ GeV$^2$ by the following expressions
($W$ in GeV, $Q^2$ in GeV$^2$): 
\begin{equation}
\sigma_{\gamma^{(*)}\gamma}(W,Q^2)=
\frac{1}{\,(Q^2+0.6)} *\left( A_0(Q^2)+A_1(Q^2)\,\log(W/20) +
A_2(Q^2)\,(\log(W/20))^2 \right)
\end{equation}
For the soft pomeron contribution we have:
\begin{eqnarray*}
A_0&=& \frac{724.6}{Q^2+5.31}\\
A_1&=& \frac{142.9}{1.666 + 3.114 \exp(-Q^2) + \sqrt{Q^2}}\\
A_3&=& 1.11 \sqrt{Q^2} +0.0311
\end{eqnarray*}
For the hard pomeron:
\begin{eqnarray*}
A_0&=& 95.67 -\frac{148.15}{Q^2+1}+\frac{79}{(Q^2+1)^2}\\
A_1&=& 68.66 - \frac{116.9}{Q^2+1}+\frac{66.6}{(Q^2+1)^2}\\
A_3&=& 31.83 - \frac{75}{Q^2+2}+\frac{11.45}{(Q^2+1)^2}\\
\end{eqnarray*}

From this expression one obtains the photon structure function for $Q^2 
\neq 0$:
\begin{equation}
\frac{1}{\alpha} \,F_2^\gamma(x,Q^2)= \frac{Q^2}{4 \pi \alpha^2} 
\sigma_{\gamma^{(*)}\gamma}(W,Q^2)
\end{equation}

The expressions for the fixed pole (box) and reggeon are given
analytically through equations (\ref{bd}) and (\ref{reggeon}) respectively.

\section{Summary}

The most significant result is that a well-tried model of diffraction
which successfully describes high-energy hadronic interactions, vector
meson production, deep inelastic scattering at small $x$, the real 
$\gamma\gamma$ cross section and the structure function of the real photon 
fails to predict correctly the $\gamma^{*}\gamma^{*}$ cross section even
at quite modest photon virtuality of $\langle Q^2 \rangle = 14.0$ GeV$^2$.
This is clearly due to the fact that, uniquely among these various processes,
the $\gamma^{*}\gamma^{*}$ interaction involves two small dipoles, and 
emphasizes the importance of the $\gamma^{*}\gamma^{*}$ cross section as 
a probe of the dynamics of the perturbative hard pomeron. If the genuinely
non-perturbative terms i.e. the soft phenomenological pomeron and Reggeon 
exchange, together with the box diagram are subtracted from the $\gamma\gamma$
and $\gamma^{*}\gamma^{*}$ data then the results can be fittedwith a single
power energy dependence $s^\epsilon$ with $\epsilon = 0.3 \pm 0.1$. The errors
on $\epsilon$ are large, partly because of the errors on current data and
partly because of the considerable uncertainty in the Reggeon term. Given 
the rather low values of $W_{\gamma^{*}\gamma^{*}}$ accessible to LEP at the 
higher values of $Q^2$ it is clearly important to get a much better 
understanding of the Reggeon term than we have at present.

We have noted that the model works very well for the real $\gamma\gamma$
cross section and for the real photon structure function due to the presence 
of two, respectively one, large dipoles. Of course the real photon is not a
hadron, with the consequence that there is an important contribution in the 
model from the hard pomeron to the real $\gamma\gamma$ cross section. However 
it has to be added that this is, as yet, not strictly required by the data.
Clarification of the remaining discrepancies between L3 and OPAL would help,
as would better data at lower energies enabling the Regge term to be more
tightly constrained. The importance of the hard pomeron is even more marked
in the case of the real photon structure function, although data are not yet 
at sufficiently small $x$ for the hard pomeron to dominate. Data for $x \le
10^{-2}$ and $Q^2 \ge 10$GeV$^2$ should clearly show its presence.

Finally we have shown that, within the model, the $\gamma^{*}\gamma^{*}$ 
cross section at small $Q_1^2$, $Q_2^2$ decreases less quickly with 
increasing $Q^2$ than is impled by naive Vector Meson Dominance. As the
model underestimates the cross section at larger $Q^2$ it is likely that
the effect at small $Q^2$ is more marked than we have indicated. As the 
real $\gamma\gamma$ cross section is obtained at present by extrapolating
from non-zero $Q^2$ using Vector Meson Dominance it is probable that it 
is significantly over-estimated.


\newpage
\begin{thebibliography}{99}

\bibitem{bfkl} V.S.Fadin, E.A.Kuraev and L.N.Lipatov: Phys.Lett. {\bf 60B}
(1975) 50\hfill\\
Y.Y.Balitskii and L.N.Lipatov: Sov.J.Nucl.Phys. {\bf 28} (1978) 822
%\hfill\\
%S. Catani and F. Hautmann: Nucl.Phys. {\bf B427} (1994) 475

\bibitem{h1}C.Adloff et al: Nucl.Phys. {\bf B497} (1997) 3

\bibitem{zeus}J.Breitweg et al: Z.Phys. {\bf C75} (1997) 215

\bibitem{fl} V.S.Fadin and L.N.Lipatov: Phys.Lett {\bf B429} (1998) 127
 
\bibitem{cc1} M.Ciafaloni and G.Camici: Phys.Lett. {\bf B430} (1998) 349

\bibitem{s} G.P.Salam: hep-ph/9806482 v2

\bibitem{cc2} M.Ciafaloni and D.Colferai: hep-ph/9812366

\bibitem{bfklp} S.J.Brodsky, V.S.Fadin, V.T.Kim, L.N.Lipatov and 
G.B. Pivovarov hep-ph/9901229

\bibitem{blm} S.J.Brodsky, G.P.Lepage and P.B.Mackenzie: Phys.Rev {\bf D28}
(1983) 228

\bibitem{dglap} V.N.Gribov and L.N.Lipatov: Sov.J.Nucl.Phys. {\bf 15}
(1972) 438 and 675\hfill\\
L.N.Lipatov: Sov.J.Nucl.Phys. {\bf 20} (1975) 94\hfill\\
Yu.L.Dokshitser: Sov.J.JETP {\bf 46} (1977) 641\hfill\\
G.Altarelli and G.Parisi: Nucl.Phys. {\bf B126} (1977) 298

\bibitem{cdl} J.R.Cudell, A.Donnachie and P.V.Landshoff: hep-ph/9901222;
Phys.Lett {\bf B}, in press

\bibitem{dl} A.Donnachie and P.V.Landshoff: Phys.Lett. {\bf B437} (1998) 408

\bibitem{t} R.Thorne: hep-ph/9901331

\bibitem{ddr} A.Donnachie, H.G.Dosch and M.Rueter: hep-ph/9810206;
Phys.Rev. {\bf D}, in press

\bibitem{bhs} S.J.Brodsky, F.Hautmann and D.E.Soper: Phys.Rev. {\bf D56}
(1997) 6957 

\bibitem{bdrw} M.Boonekamp, A.De Roeck, C.Royon and S.Wallon: hep-ph/9812523

\bibitem{bfklf} A.Bialas, R.Peschanski and C.Royon: Phys.Rev. {\bf D57} (1998)
6899\\H.Navelet, R.Peschanski and C.Royon: Nucl.Phys. {\bf B534} (1998) 297

\bibitem{BGLL67a} J.B. Bronzan, I.S.Gerstein, B.W.Lee, and F.E.Low:
Phys.Rev.Lett. {\bf 18} (1967) 32

\bibitem{BGll67b} J.B.Bronzan, I.S.Gerstein, B.W.Lee, and F.E.Low:
Phys.Rev. {\bf 157} (1967) 1448

\bibitem{Sin67} V.Singh: Phys.Rev.Lett {\bf 17} (1967) 340

\bibitem{Fub66} S.Fubini: Nuovo Cim.{\bf 43} (1966) 475

\bibitem{Fey69} R.P.Feynman: Phys.Rev.Lett. {\bf 23} (1969) 1415

\bibitem{Ter73} H.Terazawa: Rev.Mod.Phys. {\bf 45} (1973) 2105

\bibitem{Wit77} E.Witten: Nucl.Phys. {\bf B120} (1977) 189

\bibitem{HR79} C.T.Hill and G.Ross: Nucl.Phys. {\bf B148} (1979) 373

\bibitem{DGP98} H.G.Dosch, T.Gousset and H.J.Pirner: Phys. Rev. {\bf D57}
(1998) 1666

\bibitem{n} O.Nachtman: Ann.Phys. {\bf 209} (1991) 436

\bibitem{d} H.G.Dosch: Phys.Lett. {\bf B190} (1987) 555

\bibitem{ds} H.G.Dosch and Yu.A.Simonov: Phys.Lett. {\bf B205} (1988) 339

\bibitem{dfk} H.G.Dosch, E.Ferreira and A.Kr\"amer: Phys.Rev. {\bf D50} 
(1994) 1992

\bibitem{dl2} A.Donnachie and P.V.Landshoff: Phys.Lett. {\bf B296} (1992) 227

\bibitem{FP97} E. Ferreira and F. Pereira, Phys. Rev. {\bf D55} (1997) 130;
ibid {\bf D56} (1997) 179

\bibitem{BN98} E.R. Berger and O. Nachtmann, hep-ph/9808320

\bibitem{r} M.Rueter: hep-ph/9807448; Eur.Phys.J. {\bf C}, in press 

\bibitem{grv1} M.Gl\"uck, E.Reya and A.Vogt: Phys. Rev. {\bf D46} 
(1992) 1973

\bibitem{afg} P.Aurenche, M.Fontannaz and J.Ph.Guillet: Z.Phys. 
{\bf C64} (1994) 621

\bibitem{sas} G.A.Schuler and T.Sj\"ostrand: Z.Phys. {\bf C68} (1995) 607

\bibitem{aur} P.Aurenche: Phys. lett. {\bf B233} (1989) 517

\bibitem{grv2} M.Gl\"uck, E.Reya and A.Vogt: Z.Phys. {\bf C53} (1992) 651

\bibitem{pdg}
Particle Data Group: European Physical Journal {\bf 3} (1998), page 205 

\bibitem{L3a} 
L3 Collaboration: M.Acciari et al: Phys. Lett. {\bf B408} (1997) 450\\
L3 Collaboration: L3 Note 2280: Submitted to {\em XXIX ICHEP}, Vancouver, 1998

\bibitem{L399d}
L3 Collaboration: M.Acciari et al. L3 Note 2400 {\em Int. Europhys. Conf. 99}\\

\bibitem{OPALa}
OPAL Collaboration: F.W\"ackerle: {\em Proc. XXVIII Int. Symp. on 
Multiparticle Dynamics}, Frascati, 1997
\bibitem{OPA99a}
OPAL Collaboration: G. Abbiendi et al., to be submitted to Europ. Phys. J. C

\bibitem{kdp98} G.Kulzinger, H.G.Dosch and H.J.Pirner: Eur.Phys.J. {\bf C}
(to be published)

\bibitem{L3b} L3 Collaboration: M.Acciari et al: Phys.Lett. {\bf B453} (1999) 
333 and L3 Note 2404

\bibitem{L398d} L3 Collaboration: M.Acciari et al: Phys.Lett. {\bf B436}
(1998)  403 and Phys.Lett. {\bf B447} (1999) 147

\bibitem{OPA97a}
OPAL Collaboration: K.Ackerstaff et al.: Phys.Lett. {\bf B411}
(1997)  387 and Phys.Lett. {\bf B412} (1997) 225

\bibitem{OPA99b}
OPAL Collaboration: OPAL Physics Note PN389 (Preliminary)

\bibitem{ALE99a}
ALEPH Collaboration: R.Barate et al.: Phys. Lett. {\bf B458} (1999) 152

\bibitem{ALE99b}
ALEPH Collaboration: ALEPH 99-038:  EPS-HEP99 Conference, Tampere 

\end{thebibliography}
\end{document}